\documentclass[10pt,conference]{IEEEtran}
\IEEEoverridecommandlockouts

\usepackage{amsmath,amsfonts, amssymb}
\usepackage{algorithm}
\usepackage{adjustbox}
\usepackage{lscape}
\usepackage{siunitx}
\usepackage[noend]{algpseudocode}
\usepackage{textcomp}
\usepackage{fancyvrb}
\usepackage{graphicx}
\usepackage{soul}
\usepackage[many]{tcolorbox}
\usepackage{setspace}
\usepackage{booktabs}
\usepackage[utf8]{inputenc}
\usepackage[T1]{fontenc}
\usepackage{rotating}
\usepackage{graphicx}
\usepackage{paralist}
\usepackage{tabularx}
\usepackage{multicol}
\usepackage{multirow}
\usepackage{pbox}
\usepackage{enumitem}
\usepackage{colortbl}
\usepackage{pifont}
\usepackage{xspace}
\usepackage{url}
\usepackage{tikz}
\usepackage{tabularx}
\usepackage{fontawesome}
\usepackage{lscape}
\usepackage{listings}
\usepackage{color}
\usepackage{showexpl}
\usepackage{anyfontsize}
\usepackage{comment}
\usepackage{soul}
\usepackage{multibib}
\usepackage{multirow}
\usepackage{xspace}
\usepackage{footnote}
\usepackage{tcolorbox}
\usepackage{booktabs}
\usepackage{balance}
\usepackage[normalem]{ulem}

\usepackage{hhline}

\definecolor{main}{HTML}{5989cf}    % setting main color to be used
\definecolor{sub}{HTML}{cde4ff}     % setting sub color to be used

\tcbset{
	sharp corners,
	colback = white,
	before skip = 0.2cm,    % add extra space before the box
	after skip = 0.5cm      % add extra space after the box
}                           % setting global options for tcolorbox

\newtcolorbox{boxM}{
	fontupper = \color{white},
	rounded corners,
	arc = 6pt,
	colback = main!80,
	colframe = main,
	boxrule = 0pt,
	bottomrule = 4.5pt,
	enhanced,
	fuzzy shadow = {0pt}{-3pt}{-0.5pt}{0.5pt}{black!35}
}
\newcolumntype{N}{>{\centering\arraybackslash}m{.85in}}
\def\BibTeX{{\rm B\kern-.05em{\sc i\kern-.025em b}\kern-.08em
    T\kern-.1667em\lower.7ex\hbox{E}\kern-.125emX}}

\newboolean{showcomments}

\setboolean{showcomments}{true}

\ifthenelse{\boolean{showcomments}}
{\newcommand{\nb}[2]{
		\fbox{\bfseries\sffamily\scriptsize#1}
		{\sf\small$\blacktriangleright$\textit{#2}$\blacktriangleleft$}
	}
	
}
{\newcommand{\nb}[2]{}
	
}

\makeatletter
\newcommand{\linebreakand}{%
	\end{@IEEEauthorhalign}
	\hfill\mbox{}\par
	\mbox{}\hfill\begin{@IEEEauthorhalign}
}
\makeatother

\newcommand{\ie}{\emph{i.e.,}\xspace}
\newcommand{\eg}{\emph{e.g.,}\xspace}

\newcommand{\etal}{\emph{et~al.}\xspace}
\newcommand{\secref}[1]{Section~\ref{#1}\xspace}
\newcommand{\figref}[1]{Fig.~\ref{#1}\xspace}

\newcommand{\tabref}[1]{Table~\ref{#1}\xspace}

\newcommand{\java}{\emph{Java}\xspace}

\newcommand{\side}[1]{SIDE$_{{#1}}$\xspace}

\definecolor{lightergray}{rgb}{0.9,0.9,0.9}
\newtcolorbox{resultbox}{colback=lightergray, arc=0.5mm, top=2mm, bottom=2mm, left=2mm, right=2mm}

\definecolor{arsenic}{rgb}{0.23, 0.27, 0.29}
\definecolor{darkgray}{rgb}{0.33, 0.33, 0.33}

\newcommand\rev[1]{\textcolor{black}{#1}}
\begin{document}

\title{Optimizing Datasets for Code Summarization:\\ Is Code-Comment Coherence Enough?}

\author{
\IEEEauthorblockN{Antonio Vitale\IEEEauthorrefmark{1}\IEEEauthorrefmark{3}, Antonio Mastropaolo\IEEEauthorrefmark{2}, Rocco Oliveto\IEEEauthorrefmark{3}, Massimiliano Di Penta\IEEEauthorrefmark{4}, and Simone Scalabrino\IEEEauthorrefmark{4}}
\IEEEauthorblockA{\IEEEauthorrefmark{1}Politecnico di Torino, Italy, antonio.vitale@polito.it}
\IEEEauthorblockA{\IEEEauthorrefmark{2}William \& Mary, USA, amastropaolo@wm.edu}
\IEEEauthorblockA{\IEEEauthorrefmark{3}University of Molise, Italy, \{rocco.oliveto, simone.scalabrino\}@unimol.it}
\IEEEauthorblockA{\IEEEauthorrefmark{4}University of Sannio, Italy, dipenta@unisannio.it}
}

\maketitle

\thispagestyle{empty}

\begin{abstract}
Automated code summarization is a long-standing goal for code comprehension. This task automatically generates documentation using a given method. Deep Learning (DL)-based approaches have been proven beneficial for various software engineering (SE) tasks, including this one. 
Most state-of-the-art datasets for code summarization are automatically mined from GitHub and, thus, might contain erroneous or sub-optimal examples. Previous work showed that using a simple rule-based approach for removing noisy instances allows for a tangible reduction of the training set size while not reducing the effectiveness of the trained models. Motivated by this finding, we conjecture that it is possible to further reduce the dataset size by removing instances that contain different issues.
In this paper, we explore the extent to which code-comment coherence, a specific quality attribute of code summaries, can be used to optimize code summarization datasets. Specifically, we hypothesize that removing incoherent code-comment pairs might positively impact the effectiveness of the models. To do this, we rely on SIDE, a recently introduced metric for code-summary coherence. We examine multiple selectivity levels of training instances from two state-of-the-art datasets (TL-CodeSum and Funcom) and evaluate the resulting models on three manually curated test sets. The results show that even halving the training set sizes does not significantly affect the model's ability to generate summaries. However, when comparing the most restrictive selection strategy with a simpler one that randomly selects the training instances, we observe that the resulting accuracy of the model also does not change.
This result suggests that (i) current datasets contain many irrelevant examples, and (ii) different quality attributes should be explored for optimizing code summarization datasets.
\end{abstract}


\section{Introduction}
\label{sec:intro}
% !TEX root = main.tex
Writing comments and keeping them up to date during software maintenance and evolution is a challenging task and requires effort \cite{fluri2007code,fluri2009analyzing,linares2015developers,wen2019large}--yet, 
high-quality comments and documentation are essential for understanding code \cite{rani2023decade,de2005study}. Thus, automatically generating code comments from code has always been a long-lasting dream of developers and practitioners willing to bolster program comprehension in a cost-effective way \cite{haque2020improved,hu2020deep,leclair2019neural,zhang2020retrieval,mastropaolo2021studying,gao2023code}. A dream that came reality, thanks to the recent advancements in Deep Learning (DL) models and particularly Large Language Models (LLMs) that pushed the boundaries of Software Engineering (SE) automation to the next level.

Modern DL-based approaches based on Transformers \cite{vaswani2017attention,devlin2018bert,raffel2020exploring} rely on transfer learning. First, a basic model is pre-trained to acquire knowledge of the programming language. Then, several specialized models can be fine-tuned from it to tackle specific tasks.
This approach has been proven effective for several SE tasks, including code generation \cite{wei2019code,svyatkovskiy2020intellicode,liu2024your,ugare2024improving}, program repair \cite{jin2023inferfix,chen2022neural,chen2019sequencer,tufano2018empirical}, and, indeed, code summarization \cite{hu2018summarizing,hu2018deep,zhang2020retrieval,gao2023code,leclair2020improved,mastropaolo2021studying}.

Fine-tuning a DL model still requires plenty of examples, which can hardly be produced or curated manually. Thus, most state-of-the-art datasets for training DL models for coding tasks (including code summarization) are built by automatically mining open-source software repositories. 

\begin{figure}[t]
	\centering
	\includegraphics[width=\linewidth]{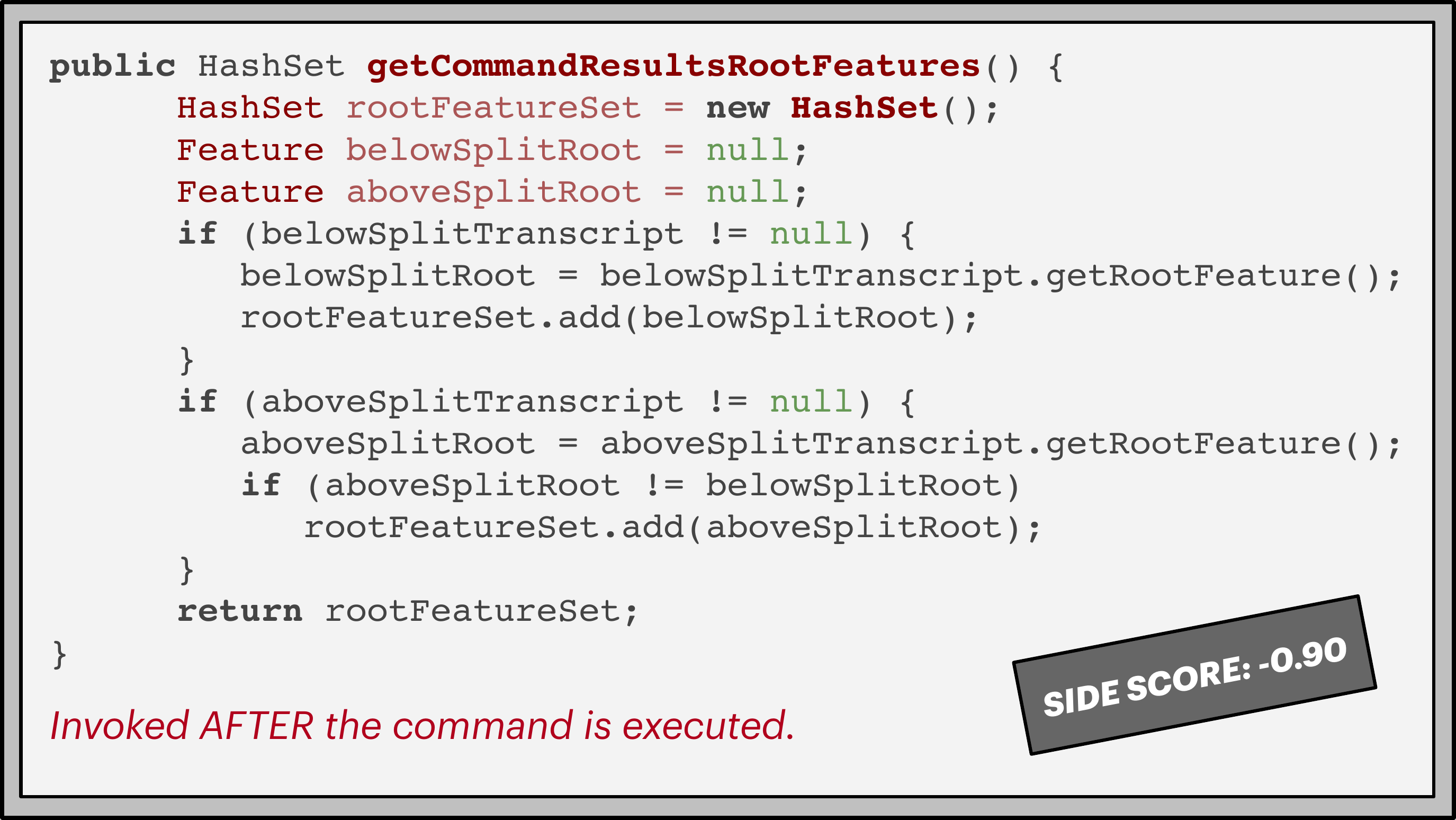}
	\caption{Example of a $\langle code, summary \rangle$ pair exhibiting a misalignment after the coarse-grained filtering by CAT.}
	\label{fig:intro-example}
\end{figure}

However, datasets created by mining software repositories tend to be noisy, containing several low-quality instances \cite{BirdBADBFD09,HerzigJZ13,BachmannBRDB10}. More specifically, Shi \etal \cite{shi2022we} recently proposed a heuristic-based approach named CAT that can automatically clean up low-quality $\langle summary, code \rangle$ pairs from code summarization datasets. This approach removes or fixes \textit{structural} quality issues at comment- (\eg commented code) and code-level (\eg empty methods). The results of their empirical study show that removing noisy instances does not reduce and even improves the model's ability to produce meaningful code summaries. 
Even though CAT streamlines coarse-grained filtering, the $\langle code, summary \rangle$ pairs that are not discarded can still have low quality and thus negatively impact the ability of the model to produce good code summaries. 
Let us consider the instance in \figref{fig:intro-example} from the state-of-the-art \textit{Funcom} dataset. While no structural problem exists (CAT does not discard it), the instance is characterized by a summary completely incoherent with the source code. Exposing the model to instances with such inconsistencies can cause hallucinations \cite{dziri2022origin}, which may lead to decreased performance.

For this reason, we conjecture that a finer-grained selection of $\langle code, summary \rangle$ pairs in code summarization datasets is not only possible, but even desirable to (i) reduce the training time, and (ii) possibly improve the model effectiveness. More specifically, we conjecture that code-comment coherence, which is a well-known quality attribute of $\langle code, summary \rangle$ pairs \cite{corazza2015coherence}, might play a crucial role in achieving this goal.

In this paper, we present an empirical investigation in which we study the impact of fine-grained filtering based on code-comment coherence on code summarization models in terms of \textit{effectiveness} (\ie correctness of the inferences) and \textit{efficiency} (\ie training time). 
To measure the code-comment coherence, we rely on SIDE, a metric recently introduced by Mastropaolo \etal \cite{mastropaolo2024evaluating}. The authors show that SIDE strongly correlates with human evaluations of summary quality, surpassing established metrics including BLEU \cite{papineni2002bleu} and ROUGE \cite{lin2004rouge}.
We consider two state-of-the-art datasets for code summarization, \ie TL-CodeSum \cite{hu2018summarizing} and Funcom \cite{leclair2019neural}, already filtered with CAT \cite{shi2022we}. Then, we further filter the instances in terms of SIDE value by considering different thresholds (\ie $\{0.5, 0.6, 0.7, 0.8, 0.9\}$). Finally, we use each resulting training set (including the original one) to fine-tune \emph{CodeT5+} \cite{wang2023codet5+}.

We test the models on two manually-curated datasets from the literature \cite{yu2024codereval, mastropaolo2023robustness}. We observe that reducing the size of the training set, even with the most restrictive filter (SIDE$_{0.9}$, which selects $\sim$50\% of the instances), has a negligible impact on the model's effectiveness. On the other hand, reducing the number of training instances results in a significantly lower training time (up to $\sim$111 saved hours).
To further validate our original hypothesis that code-comment coherence is a suitable quality attribute for filtering instances in code summarization datasets, we compared the most restrictive filter (SIDE$_{0.9}$) with a filter that keeps the same number of instances, but by simply choosing them randomly. Surprisingly, we observed that the random filter achieves negligibly worse results than SIDE$_{0.9}$.

Our results provide two clear insights. First, code-comment coherence is marginally important for selecting suitable instances for code summarization. Second, regardless of this, removing instances (even randomly!) does not impact the effectiveness of code summarization models.
This suggests that additional code-comment quality attributes should be investigated, and researchers should prioritize relevance over quantity to better select the most informative instances to build code summarization datasets.

The paper is organized as follows. Section \ref{sec:selection_strategies} provides backgrounds on the existing selection technique for code summarization (CAT \cite{shi2022we}) and overviews SIDE \cite{mastropaolo2024evaluating}. Section \ref{sec:study} details the study definition and planning. Results are reported in Section \ref{sec:result}, while Section \ref{sec:implication} discusses its implications, and Section \ref{sec:threats} the threats to its validity. Section \ref{sec:related} discusses related work about data quality. Finally, Section \ref{sec:conclusion} concludes the paper and outlines directions for future work.

\section{Selection Strategies for Code Summarization}
\label{sec:selection_strategies}
% !TEX root = main.tex
In this section, we provide backgrounds about (i) a state-of-the-art strategy for repairing or removing poor instances from code summarization datasets (CAT), and (ii) the SIDE metric, which we use to streamline a data-centric, quality-aware instance filtering.

\subsection{Code-comment cleAning Tool}
CAT (\textbf{C}ode-comment cle\textbf{A}ning \textbf{T}ool) is an approach and tool by Shi \etal \cite{shi2022we} to detect and handle noisy instances given the pairs of \textit{<code, summary>} from code-summarization datasets. The development of CAT was preceded by a manual investigation involving 9 participants who examined 1,600 \textit{<code, summary>} pairs. This manual analysis aimed to define a taxonomy of noisy data categories.

The taxonomy features comments- and code-related noisy data, such as \textit{commented-out method} or \textit{empty function}. Based on such categories, Shi \etal defined a set of heuristic rules and implemented them in the CAT tool. CAT works with two possible strategies based on the issues found: On the one hand, it fixes instances with minor issues. For example, it removes block-level comments. On the other hand, it completely drops instances where no fix is possible, including, for example, getters and setters.

\subsection{Fine-Grained Filtering: SIDE}
\label{sec:side-aware}
SIDE (\textbf{S}ummary al\textbf{I}gnment to co\textbf{D}e s\textbf{E}mantics) is a novel quality-aware metric presented by Mastropaolo \etal \cite{mastropaolo2024evaluating}. SIDE addresses the shortcomings of traditional metrics such as BLEU \cite{papineni2002bleu}, ROUGE \cite{lin:tsbo2004}, and METEOR \cite{banerjee:acl2005} used in code summarization tasks. 

SIDE employs a contrastive learning approach to determine the accuracy with which a code summary documents the underlying code, explicitly focusing on Java methods. Contrastive learning aims to maximize the distance between the reference code and inappropriate comments while minimizing the distance to suitable comments. The model that implements SIDE, MPNet \cite{Song2020MPNetMA}, provides a continuous score ranging from -1 to 1. Scores closer to~-1 indicate poor alignment between the code summary and the actual code, whereas scores closer to~1 suggest a strong alignment. SIDE showed a high correlation with human evaluations of summary quality, outperforming established metrics like BLEU, ROUGE, and METEOR. 

The benefits of a quality-aware metric like SIDE extend beyond evaluating code summarization techniques, as Mastropaolo \etal \cite{mastropaolo2024evaluating} noted. SIDE can be valuable when distinguishing high-quality code documentation from subpar examples is essential. We conjecture that SIDE ensures that only the instances most likely to enhance the training procedure are used, enabling the model to converge faster without a drop in performance by filtering out low-quality elements.

Our study relies on SIDE to select $\langle code, summary \rangle$ pairs with high coherence. Specifically, given a training set $T$ and a threshold $t$, we select the training instances $\{ p_{i} \in T \mid \text{SIDE}(p_{i}) \ge t \}$.

\section{Study Definition, Design and Planning}
\label{sec:study}
% !TEX root = main.tex
The \emph{goal} of this study is to empirically evaluate how code-comment coherence, through a quality-aware selection strategy grounded on SIDE, impacts the effectiveness and training efficiency of neural code summarization models.

More specifically, the study aims to address the following research questions:

\begin{itemize}[itemindent=0.25cm]
	\item[\textbf{RQ$_{0}$}:] \textit{How do code summarization datasets measure up in terms of code-comment coherence?}
	In this preliminary question, we assess the coherence of code-comment pairs of datasets commonly used in code summarization. As we aim to use a coherence-aware strategy to optimize training sets, first of all, we would like to see how the coherence is distributed.
	\item[\textbf{RQ$_{1}$}:] \textit{How does a coherence-aware strategy selection impact the performance of neural code summarization models?}
	In this research question, we investigate how a targeted selection of training data based on code-comment coherence impacts the performance of neural code summarization models.
	\item[\textbf{RQ$_{2}$}:] \textit{How does the coherence-aware strategy selection compare with a random baseline?}
	In this research question, we test our hypothesis that code-comment coherence is a quality attribute that can be used to select training instances.
\end{itemize}

\subsection{Context Selection}
\label{subsec:context_selection}
The \emph{context} of our study consists of datasets containing pairs of \java methods with the associated summaries. 
For fine-tuning the models, we consider the two most important datasets from the state of the art: TL-CodeSum \cite{hu2018summarizing}, and Funcom \cite{leclair2019neural}.

The \textit{TL-CodeSum} dataset \cite{hu2018summarizing} is specifically designed for the code summarization task. It consists of $\sim$87k instances $ \langle code, summary \rangle$ extracted from GitHub repositories created from 2015 to 2016, and having at least 20 stars. In detail, Hu \etal \cite{hu2018summarizing} extracted the first sentence---likely to describe the overall method functionality---from the doc of each pair.

Similarly to TL-CodeSum, the \textit{Funcom} dataset \cite{leclair2019neural} is also specifically designed for code summarization. \textit{Funcom} consists of over 2.1M $ \langle code, summary \rangle$ pairs collected from the Sourcerer repository. As for TL-CodeSum, LeClair \etal \cite{leclair2019neural} only consider methods with their javadoc, extracting the first sentence as corresponding \textit{summary}.

Shi \etal \cite{shi2022we} found many noisy instances and duplicates in the above-described datasets and cleaned them up using their heuristic-based dataset-cleaning approach. For this reason, we use the cleaned versions of \textit{TL-CodeSum} and \textit{Funcom} provided by Shi \etal \cite{shi2022we}. The cleaned \textit{TL-CodeSum} contains 53,597 training instances, while the cleaned \textit{Funcom} contains 1,184,438 training instances.

The above datasets are built automatically, and no manual check was performed, \ie there is no guarantee of their quality. For this reason, we use two additional, manually curated datasets to test the models. The first one is \textit{CoderEval} \cite{yu2024codereval}, which consists of 230 Python and 230 \java code generation problems collected from open-source, high-starred projects which include \textit{original} and \textit{human-labeled} docstrings that should act as prompt for Code Generation models to generate the corresponding \textit{code}. 
The instances have been subject to manual screening, for which the main criterion is the probability of appearing in real-development scenarios. We focus on the \java set of problems, inverting the input and the output \ie from $\langle docstring, code \rangle$ to $\langle code, docstring \rangle$. 
To align the format of the pairs format, we performed an additional manual analysis in which one of the authors checked all the triplets with a second author to confirm the analysis. 
We found that some of the \textit{docstring}(s) contained more than a sentence. Therefore, to make them consistent with the previous dataset format (\eg single sentence), we extracted the first sentence from each \textit{docstring}. Still, we found 12 occurrences in which the corresponding \textit{original docstring} does not describe the \textit{code} (\eg ``\texttt{{@inheritDoc}}'', ``\texttt{@param modelName model name of the entity}'', and similar). We also excluded \textit{docstring}: ``\texttt{Computes floor(\$log\_2 (n)\$) \$+ 1\$.}'' since it includes a formula not explained in natural language.
Again, to appropriately align the evaluation, we do not evaluate such instances, ending up with 218 \textit{original} instances.

The second manually-curated dataset we use is the one by Mastropaolo \etal \cite{mastropaolo2023robustness}. The dataset consists of 892 methods associated with their summary (\ie first sentence of the method documentation), collected from non-fork GitHub \java repositories with at least 300 commits, 50 contributors, and 25 stars. 
Such instances are in the form $\langle summary, code\rangle$ and, as for CoderEval \cite{yu2024codereval}, we inverted the input and the output \ie $\langle code, summary\rangle$. Mastropaolo \etal  analyzed such pairs to ensure their quality. We manually analyzed and cleaned them further (\eg ``\texttt{Adds an {@link CarrierService} to the {@linkCarrier}}'' into ``\texttt{Adds an CarrierService to the Carrier}''), as we had done for CoderEval. No instances were removed during such a manual analysis.

We remove the instances from the test sets which appear in the training sets of \textit{TL-CodeSum} and \textit{Funcom}. As a result, we remove ten instances from CoderEval, which are present only in the \textit{TL-CodeSum} training set.

\subsection{Study Methodology}
\label{subsec:exp_proc}
To answer RQ$_{0}$, we use SIDE to compute the degree to which the summaries of the studied datasets document their corresponding code. We did this for each instance of the training sets included in \textit{TL-CodeSum} and \textit{Funcom}. To understand the coherence of the training sets, we analyze the average and the distributions of the SIDE scores of the instances.\\

\addtolength{\extrarowheight}{\belowrulesep}
\aboverulesep=0pt
\belowrulesep=0pt
\begin{table}[t]
	\centering
	\caption{Different selections for \textit{TL-CodeSum} and \textit{Funcom} training sets.}
	\label{tab:dataset_w_strategies}
	\resizebox{0.6\columnwidth}{!}{%
		\begin{tabular}{lrrr}
			\toprule
			\cellcolor{black}\textcolor{white}{\textbf{Selection}} &  \cellcolor{black}\textcolor{white}{\textbf{TL-CodeSum}} & \cellcolor{black}\textcolor{white}{\textbf{Funcom}} \\
			\midrule
			Full & 53,597 & 1,184,438 \\
			\midrule
			\side{0.5} & 50,073 & 1,080,649 \\
			\side{0.6} & 48,146 & 1,031,647 \\
			\side{0.7} & 44,853 & 952,265 \\
			\side{0.8} & 38,733 & 813,998 \\
			\side{0.9} & 26,258 & 540,170 \\
			\bottomrule
	\end{tabular}}
\end{table}

To answer RQ$_{1}$, we use the SIDE-based filter we define in \secref{sec:selection_strategies}. We use five threshold values, \ie 0.5, 0.6, 0.7, 0.8, and 0.9. We do not use thresholds lower than 0.5 because they would result in negligible dataset reductions (lower than 10\% for both), as we will observe in the results of RQ$_{0}$.
We report information about the different datasets in \tabref{tab:dataset_w_strategies}. 
We apply each filter on the training sets of \textit{TL-CodeSum} and \textit{Funcom}. Such filtering leads to the definition of five new versions of both datasets.

We fine-tune a pre-trained Transformer-based model for each dataset version, \ie both the base one and its six filtered versions, producing 12 fine-tuned models.
We choose to leverage the pre-trained \emph{CodeT5+} \cite{wang2023codet5+} since it has been largely used for code-related tasks \cite{ahmed2024automatic,phan2024repohyper,yang2024important} and, more important, in the code summarization approaches described above. This model is built on the backbone of the well-known T5 model by Raffel \etal \cite{raffel2020exploring}, yet it benefits from specific enhancements tailored for code understanding and generation tasks. During the pre-training phase, \emph{CodeT5+} is first trained on unimodal data, which includes code and comments, employing a combination of pre-training objectives such as span-denoising \cite{raffel2020exploring} and Causal Language Modeling \cite{soltan2022alexatm,tay2022ul2}. Then, it is pre-trained on bi-modal data where pre-training objectives such as text-code contrastive learning, text-code matching, and text-code causal language modeling are employed. It comes with different variants: (i) \emph{CodeT5+} 220M, (ii) \emph{CodeT5+} 770M, (iii) \emph{CodeT5+} 2B, (iv) \emph{CodeT5+} 6B, and (vi) \emph{CodeT5+} 16B.
Since our experimental design would require training, validating and testing 12 models, we decided to fine-tune the \emph{CodeT5+} variant featuring 220M trainable parameters.
This choice aligns with the goal of our investigation: Rather than proposing a new code summarization technique, we aim to use a model that offers a favorable balance between size and training time while still allowing us to observe the relevant phenomenon (if present).

Considering the extensive array of our experiments, we fine-tune for 20 epochs using a batch size of 16. Additionally, we restrict the input length to 512 tokens and the output to 128 tokens, consistent with previous studies leveraging the two datasets we used \cite{mastropaolo2022using,zhou2022automatic,tufano2023automating}. In addition, we conduct the fine-tuning using the standard hyperparameters for \emph{CodeT5+}, which include the AdamW optimizer \cite{loshchilov2017decoupled} and a learning rate of 2e-5, which is the one recommended for (Code)T5 and also used in works leveraging such models \cite{mastropaolo2023towards,ciniselli2024generalizability,mastropaolo2024vul}.

To prevent overfitting, we employ early stopping \cite{prechelt2002early}. After each epoch, we assess the performance of the models by computing the number of correct predictions on the validation set. 
In line with similar research \cite{mastropaolo2023towards,ciniselli2024generalizability}, we implement early stopping with patience of 5 epochs and a delta of 0.01. This means that training will stop if the model's performance does not improve by at least 0.01 for five consecutive epochs. We then select the best-performing checkpoint before early stopping.
We fine-tune a \emph{CodeT5+} model for each training set derived from the selection strategy \ie 12 (2 datasets $\times$ six variants).

After training the models, we assess their performance on the test set dataset that, as previously explained, are the \textit{CoderEval} \cite{yu2024codereval}, and the one from Mastropaolo \etal \cite{mastropaolo2023robustness} which we refer to as the \textit{golden sets}.

In the inference phase, we employ a beam search decoding strategy. In detail, with $k \in \{1, 3, 5\}$, we allow each model to generate the $k$ most probable candidate \textit{summaries} for the given \textit{code}.
To evaluate the generated summaries of each model, we compute the following metrics: BLEU \cite{papineni2002bleu}, METEOR \cite{banerjee:acl2005}, and ROUGE-L \cite{lin2004rouge}.
\textbf{BLEU} is a metric that expresses, within a range from 0 to 1, the similarity between a generated text (candidate) and the target one (oracle). It computes the percentage of $n$-grams of the generated text that appear in the target, where $n \in \{1, 2, 3, 4\}$. 
\textbf{METEOR} is computed as the harmonic mean of unigram precision and recall, with the latter weighted higher than the former. It ranges from 0 to 1. 
\textbf{ROUGE-L} is computed as the length of the longest common subsequence (LCS) between the generated text and the target one and measures the recall by considering the proportion of the LCS relative to the length of the target text.
We do not use SIDE \cite{mastropaolo2024evaluating} as it was employed for selecting training instances and could therefore be unnaturally biased in favor of models trained on filtered datasets.
Also, we do not compute the percentage of exact matches for three reasons. First, exact matches might underestimate the actual performances of the model. Indeed, an exact match implies a correct summary, but many alternative summaries might be as correct (or even more correct, in theory) as the ones in the ground truth for the very nature of this task. Second (also related to the previous point), \textit{CoderEval} \cite{yu2024codereval} provides two summaries for each coding instance, namely \textit{original} (\ie the docstring collected from the original source code), and \textit{human} (\ie the docstring written from scratch by developers during the benchmark creation \cite{yu2024codereval}). The model could have correctly generated only one of them, which are, by definition, both correct alternatives, thus leading to inconsistent results. Third, the dataset provided by Mastropaolo \etal~\cite{mastropaolo2023robustness} includes three different yet semantically equivalent code summaries for each \java method. As previously noted, each of these alternative descriptions is a valid candidate summary.

\begin{figure}[t]
	\centering
	\includegraphics[width=0.65\linewidth]{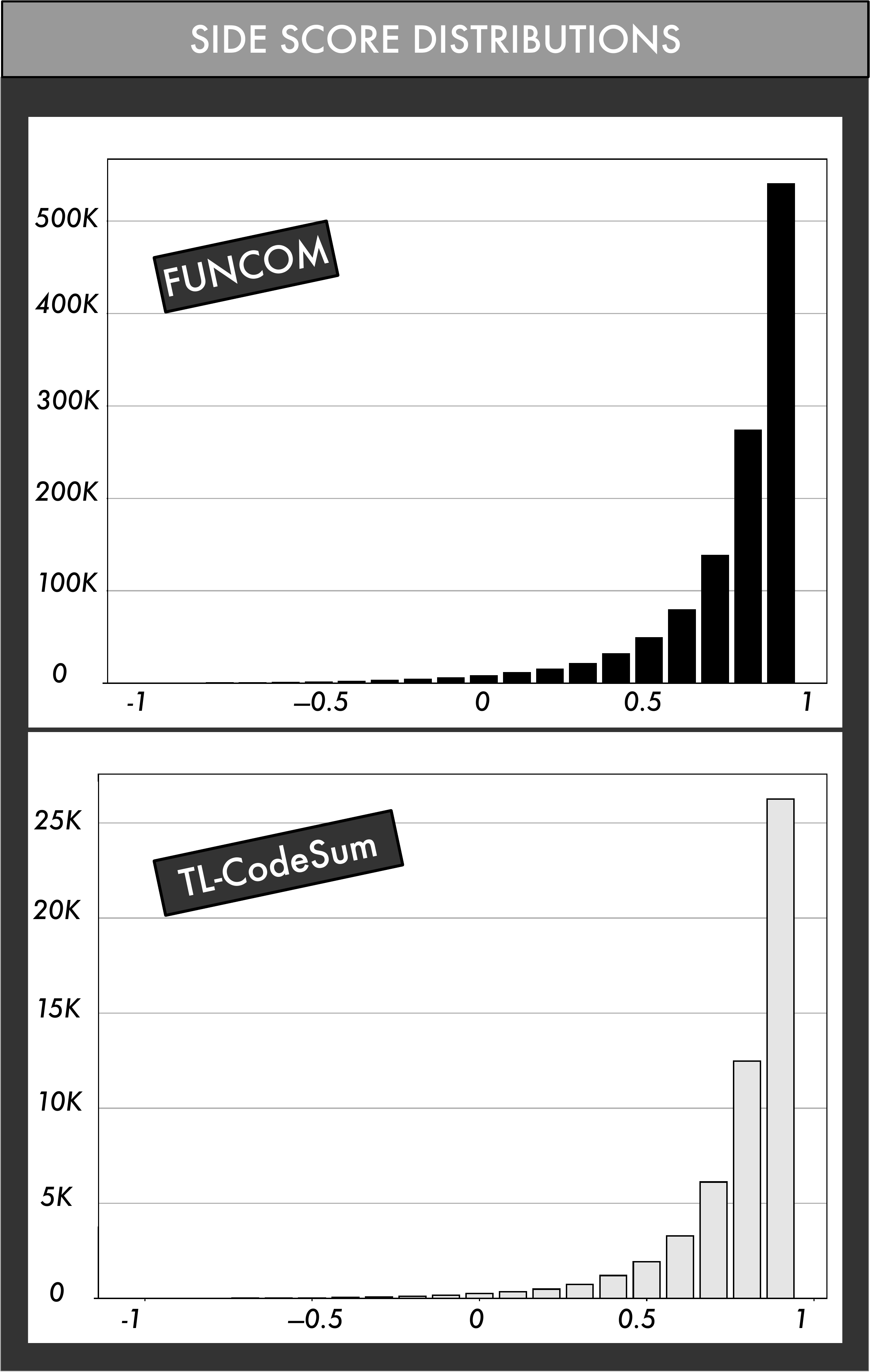}
	\caption{Distribution of SIDE scores for \textit{TL-CodeSum} and \textit{Funcom} training instances.}
	\label{fig:rq0_training_distribution}
\end{figure}

\begin{table*}[t]
	\centering
	\caption{Performance metrics on Top-1 predictions for CoderEval.}
	\label{tab:performance_metrics_codereval}
	\resizebox{\linewidth}{!}{%
		\begin{tabular}{c|l|r|r|rrr|rrr|rrr}
			\toprule
			\rowcolor{black}
			&  &  &  & \multicolumn{3}{c}{\textcolor{white}{\textbf{CoderEval-Original \cite{yu2024codereval}}}} & \multicolumn{3}{c}{\textcolor{white}{\textbf{CoderEval-Human \cite{yu2024codereval}}}} & \multicolumn{3}{c}{\textcolor{white}{\textbf{Mastropaolo \etal \cite{mastropaolo2023robustness}}}} \\
			\rowcolor{gray!20}
			\textbf{Dataset} & \textbf{Selection} & \textbf{\#Tokens} & \textbf{(\%) Saving} & \textbf{BLEU-4} & \textbf{METEOR} & \textbf{ROUGE} & \textbf{BLEU-4} & \textbf{METEOR} & \textbf{ROUGE} & \textbf{BLEU-4} & \textbf{METEOR} & \textbf{ROUGE} \\
			\midrule
			\multirow{6}{*}{\textit{TL-CodeSum \cite{hu2018summarizing}}}
			
			& \emph{Full}      & \cellcolor[HTML]{656565}\color[HTML]{FFFFFF} $\uparrow$ 7.9M &  \cellcolor[HTML]{a3070c}\color[HTML]{FFFFFF} --   & 11.72 & 17.84 & 35.01 & 6.41 & 14.28 & 30.51 & 6.37 & 13.04 & 27.09 \\
			\cmidrule(r){2-13}
			& \side{0.5} & 7.4M & 7\%  & 12.41 & 17.73 & 35.41 & 6.32 & 14.28 & 31.08 & 6.36 & 12.93 & 27.47\\
			& \side{0.6} & 7.1M & 10\% & 13.22 & 17.98 & 36.49 & 6.98 & 14.03 & 31.06 & 6.61 & 13.07 & 27.49\\
			& \side{0.7} & 6.6M & 16\% & 13.17 & 17.93 & 36.14 & 6.79 & 14.83 & 31.90 & 6.30 & 12.95 & 27.60\\
			& \side{0.8} & 5.6M & 28\% & 11.99 & 17.54 & 34.60 & 6.23 & 13.94 & 29.72 & 6.02 & 13.01 & 27.63\\
			& \side{0.9} & \cellcolor[HTML]{656565}\color[HTML]{FFFFFF}  $\downarrow$ 3.7M & \cellcolor[HTML]{026329}\color[HTML]{FFFFFF} 51\% & 11.60 & 17.07 & 33.67 & 6.56 & 14.19 & 30.53 & 5.62 & 12.75 & 27.26   \\
			\midrule
			\rowcolor[gray]{.85} & & & & & & & & & & & & \\
			\midrule
			\multirow{6}{*}{\textit{Funcom \cite{leclair2019neural}}} & \emph{Full}      & \cellcolor[HTML]{656565}\color[HTML]{FFFFFF} $\uparrow$ 108.7M  & \cellcolor[HTML]{a3070c}\color[HTML]{FFFFFF} --   & 14.25 & 17.61 & 36.53 & 5.93 & 12.95 & 28.36 & 6.77 & 12.94 & 28.00 \\
			\cmidrule(r){2-13}
			& \side{0.5} & 99.5M & 9\%  & 15.04 & 18.16 & 37.08 & 6.15 & 13.14 & 29.05 & 6.84 & 12.87 & 27.93 \\
			& \side{0.6} & 95.0M & 13\% & 16.19 & 19.30 & 38.38 & 7.02 & 14.07 & 30.25 & 7.03 & 12.99 & 28.33 \\
			& \side{0.7} & 87.7M & 20\% & 14.77 & 18.92 & 37.39 & 7.49 & 14.19 & 30.14 & 6.62 & 12.79 & 27.78 \\
			& \side{0.8} & 74.2M & 31\% & 14.10 & 18.34 & 37.04 & 6.53 & 13.38 & 28.45 & 6.93 & 12.78 & 27.99 \\
			& \side{0.9} & \cellcolor[HTML]{656565}\color[HTML]{FFFFFF} $\downarrow$ 49.0M & \cellcolor[HTML]{026329}\color[HTML]{FFFFFF} 54\% & 13.65 & 18.08 & 37.12 & 6.78 & 13.61 & 30.07 & 6.81 & 12.95 & 28.07 \\
			\bottomrule
	\end{tabular}}
\end{table*}

We also perform statistical hypothesis tests (Wilcoxon signed-rank test) \cite{wilcoxon1992individual} and Cliff's delta effect size \cite{grissom2005effect} to compare the distributions of the BLEU-4, METEOR, and ROUGE-L of the predictions generated by the different models trained on the filtered training sets with those of the models trained on the full training sets. We use Holm's correction \cite{holm1979simple} to adjust the \textit{p}-values for the multiple tests. We reject the \textit{null hypothesis} (there is no difference between the effectiveness of two given models) if the \emph{p}-value is lower than 0.05.

Finally, we study the Pareto front to analyze the cost-benefit trade-offs between the effectiveness of the models trained on the different selections of \textit{TL-CodeSum} and \textit{Funcom} (benefit, measured with \ie, BLEU, METEOR, and ROUGE-L) and the corresponding training dataset size (cost).

To answer RQ$_{2}$ we compare the selection strategy with \side{0.9} (\ie the most restrictive selection), with a \textit{Random} baseline. In detail, we randomly sample the same number of training instances as those selected with \side{0.9} from each dataset. We compare the effectiveness of the models trained with the training instances selected with \side{0.9} and \textit{Random} measured in terms of the previously described metrics (\ie BLEU-4, METEOR, and ROUGE-L). Again, we perform statistical hypothesis tests (Wilcoxon signed-rank test) \cite{wilcoxon1992individual} and compute the Cliff's delta effect size \cite{grissom2005effect} to compare the distributions of BLEU-4, METEOR, and ROUGE-L of the predictions generated by the \side{0.9} model and the \textit{Random} baseline. We use Holm's correction \cite{holm1979simple} to adjust the \textit{p}-values for the multiple tests. We reject the \textit{null hypothesis} (there is no difference between the two models) if the \emph{p}-value is lower than 0.05.

\section{Results}
\label{sec:result}
In this section, we present the outcomes of our study, addressing the research questions formulated in Section \ref{sec:study}.

\subsection{RQ$_{0}$ Consistency Analysis}
\label{rq0}
\figref{fig:rq0_training_distribution} reports the SIDE score distributions for the training instances of the examined datasets. We can observe that most of the instances of the two datasets exhibit high-quality features, according to SIDE. In detail, the mean SIDE scores for the training sets are 0.83 and 0.81 for \textit{TL-CodeSum} and \textit{Funcom}, respectively. This result denotes a high coherence in the $\langle code, summary \rangle$ pairs. Also, we can notice that, for more than 90\% of the instances (93\% for \textit{TL-CodeSum} and 91\% for \textit{Funcom}), the SIDE score is greater than 0.5. While such a percentage is quite high, it is worth noting that, as a consequence, a large number of instances (3,525 and 103,789 for \textit{TL-CodeSum} and \textit{Funcom}, respectively) exhibit a SIDE score lower than 0.5. Besides, 51\% and 54\% of the instances from \textit{TL-CodeSum} and \textit{Funcom}, respectively, have a SIDE score lower than 0.9. While many of such instances are not necessarily detrimental to the model, they might not be beneficial either. Such distributions show a sufficiently high margin of improvement for both datasets.

\begin{tcolorbox}[
	colframe=black!100!white, colback=black!10!white,
	coltitle=white, colbacktitle=black!,
	title=Answer to RQ$_{0}$,
	fonttitle=\bfseries,
	rounded corners,
	boxrule=0.5mm,
	width=\linewidth,
	breakable
	]
	
	The training sets from both \textit{TL-CodeSum} and \textit{Funcom} exhibit a high SIDE score, on average (> 0.8 for both). Still, over half the instances have a sub-optimal SIDE score (< 0.9). This distribution suggests that a dataset reduction approach could  improve the performance of an automated code summarization model trained on it.
\end{tcolorbox}

\begin{figure*}[t]
	\centering
	\includegraphics[width=0.9\textwidth]{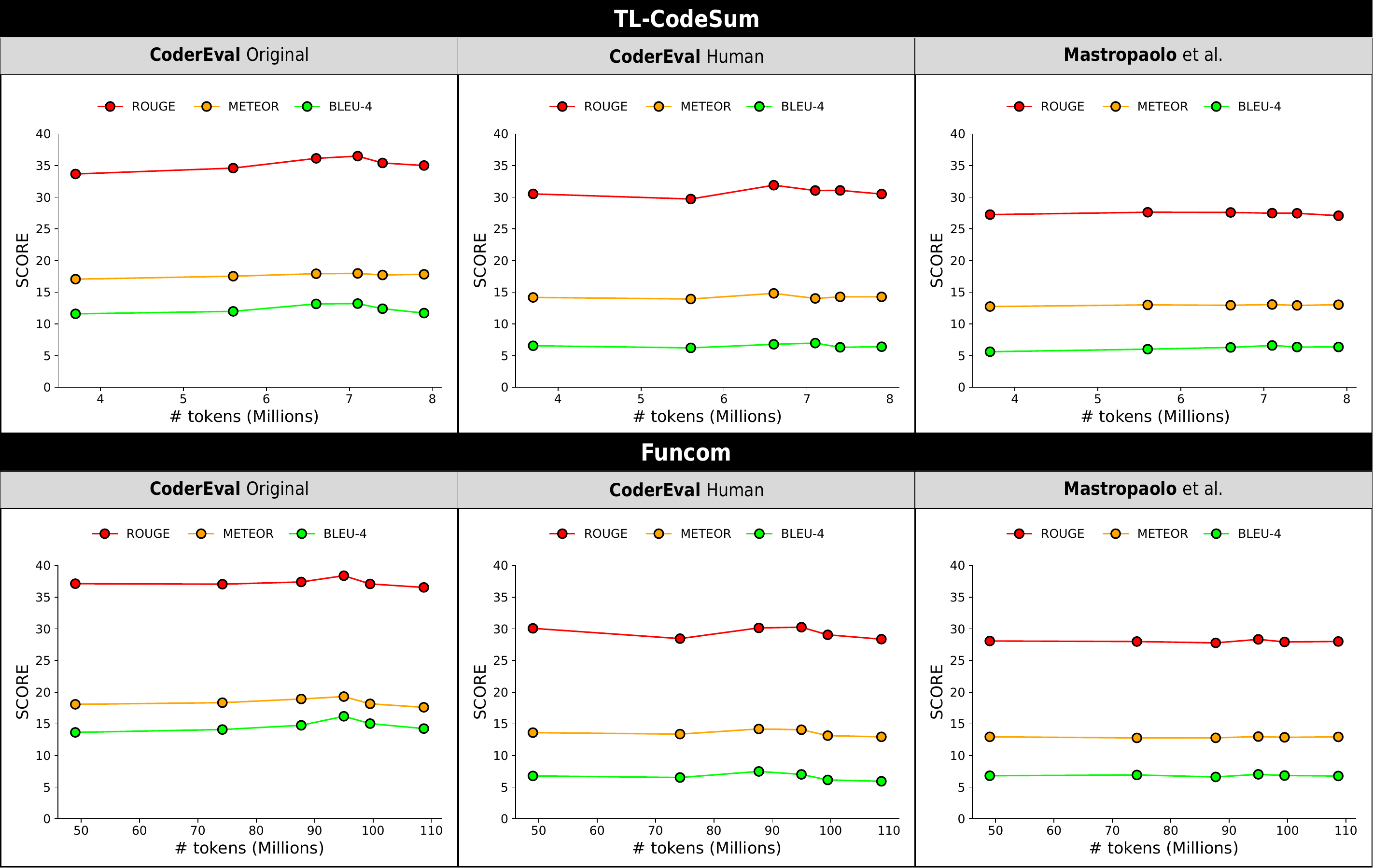}
	\caption{Pareto front for TL-CodeSum \cite{hu2018summarizing} (\textbf{top}) and Funcom \cite{leclair2019recommendations} (\textbf{bottom}). From left to right, the chart shows the number of tokens for $SIDE_{0.9}$, $SIDE_{0.8}$, $SIDE_{0.7}$, $SIDE_{0.6}$, $SIDE_{0.5}$, and the full dataset.}
	\label{fig:pareto}
\end{figure*}

\subsection{RQ$_{1}$. Selection Strategy Impact}
\label{rq1}
\tabref{tab:performance_metrics_codereval} reports the results obtained from the different models fine-tuned on the full and filtered training sets, while the detailed results of the Wilcoxon signed-rank tests are in our replication package for space reasons \cite{replicationpackage}. The column ``Dataset'' indicates the training set from which the selections were performed, while the ``Selection'' column reports the SIDE threshold used to filter the training sets \ie \textit{full}, and \side{0.5} up to \side{0.9}. For example, the first line in correspondence of \textit{TL-CodeSum} (\ie Full) represents the model fine-tuned on the full training set of \textit{TL-CodeSum}, while the one below \ie (\side{0.5}) represents the model fine-tuned on the training set filtered with SIDE score greater than 0.5.
Furthermore, we report the number of tokens seen during the fine-tuning process (``Tokens'' column) and the percentage of the saved training instances with respect to the complete training sets (``Saving'' column).
Finally, the columns ``BLEU-4'', ``METEOR'', and ``ROUGE'' report the performance in terms of percentage for the golden sets including \textit{CoderEval-Original}, \textit{CoderEval-Human}, and the one by Mastropaolo \etal~\cite{mastropaolo2023robustness}.

In the following, we discuss results achieved through the SIDE-based selection by comparing them against the \textit{CoderEval-Original} ground truth.
Looking at the rows corresponding to \textit{TL-CodeSum}, we observe that the model trained with the complete training set achieves 11.72\% BLEU-4, 17.84\% METEOR, and 35.01\% ROUGE-L. Surprisingly, even though such a model underwent fine-tuning with the full training set, it is not the one exhibiting the best results in terms of these metrics. Instead, the model fine-tuned with the selection strategy with \side{0.6} is the one that achieves the best results for all metrics despite being trained with $\sim$1M fewer tokens during training than the one leveraging the whole corpus of code tokens. If we look at the ``minimal'' selection strategy \ie \side{0.9}, we observe a limited drop in performance ($\downarrow$ 0.12, $\downarrow$ 0.77, and $\downarrow$ 1.34) while saving the 51\% of the training instances.

Looking at the rows corresponding to \textit{Funcom}, we observe the same trend as for \textit{TL-CodeSum}. The model fine-tuned on the whole training set (108.7M tokens) performs worse than the models fine-tuned with less training data. We can notice a negligible drop in performance for \side{0.9} (49.0M tokens) on BLEU-4 metric ($\downarrow$ 0.60), while achieving slightly better performance on METEOR ($\uparrow$ 0.47) and ROUGE-L ($\uparrow$ 0.59).

The Wilcoxon signed-rank tests indicate that there is no statistically significant difference between the results obtained by the model fine-tuned on the full training set and those fine-tuned on the filtered training sets (minimum \textit{p}-value~$> 0.2$ for \textit{TL-CodeSum}, and \textit{p}-value~$> 0.08$ for \textit{Funcom}). Also, the Cliff's $\delta$ effect size is always negligible for the same comparisons. Such results occur despite the most aggressive selection only leveraging 46\% of the original training instances. In other words, there is little to lose in terms of summary quality when the model training set is heavily optimized.

Similar observations apply to the \textit{CoderEval-Human} dataset. With no statistically significant differences, (minimum \textit{p}-value $> 0.23$ for \textit{TL-CodeSum}, and \textit{p}-value $> 0.15$ for \textit{Funcom}), according to the Wilcoxon signed-rank test and a negligible Cliff's $\delta$ effect size, the models specialized to produce meaningful code summaries with the filtered dataset have comparable performance to those fine-tuned with the original dataset.

For example, the model instructed with ``high-quality'' examples included in the \side{0.9} dataset --- even if trained with 4.2M fewer tokens --- obtains a slightly higher BLEU-4 ($\uparrow$ 0.15) and a slightly lower METEOR ($\downarrow$ 0.09) and ROUGE-L ($\uparrow$ 0.02). We also observed no substantial differences in the context of \textit{Funcom}. In this case, however, the models fine-tuned with the filtered training sets perform slightly better than the model fine-tuned on the full-training set in terms of all the metrics ($\uparrow$ 0.85 BLEU-4, $\uparrow$ 0.66 METEOR, and $\uparrow$ 1.71 ROUGE-L).

Finally, the conclusions above (no statistically significant differences, negligible Cliff's $\delta$ effect sizes) are further confirmed on the dataset by Mastropaolo \etal \cite{mastropaolo2023robustness}.
The model that underwent fine-tuning with the full-sized \textit{TL-CodeSum} training set performs only slightly better than the models fine-tuned on the filtered datasets. At the same time, it is interesting to notice how---although differences are still not statistically significant---the model fine-tuned on the filtered \textit{Funcom} training set with \side{0.6} exhibits marginally better performance compared to the model that was exposed to the largest amount of code tokens, \ie the full-sized \textit{Funcom} dataset, during training. This suggests that there may be contexts for which the selection based on the SIDE metric helps to prune out instances worsening the summarization quality.

\figref{fig:pareto} depicts the Pareto fronts for the models' performance and training dataset size across the different selections of the \textit{TL-CodeSum} and \textit{Funcom} datasets (\ie \textit{Full}, SIDE$_{0.5}$, SIDE$_{0.6}$, SIDE$_{0.7}$, SIDE$_{0.8}$, SIDE$_{0.9}$), evaluated on the three benchmarks \ie \textit{CoderEval-Original}, \textit{CoderEval-Human}, and Mastropaolo \etal.

For the models trained on the \textit{TL-CodeSum} selections, we observe that as the training dataset size increases from SIDE$_{0.9}$ with 3.7M tokens to SIDE$_{0.7}$ with 6.6M tokens, the ROUGE scores remain almost constant. A slight improvement in ROUGE scores can be observed at 7.1M tokens (SIDE$_{0.6}$) on \textit{CoderEval-Original} and \textit{CoderEval-Human}. However, at larger sizes with 7.4M tokens (SIDE$_{0.5}$) and the 7.9M tokens (\textit{Full}) the scores decline. At the same time, the METEOR and BLEU-4 scores remain almost constant across all dataset sizes, showing minimal improvements. In the Mastropaolo \etal dataset, the three metrics remain constant across all filtering levels, evidencing that increasing dataset size does not meaningfully impact the models' effectiveness.

The models trained on \textit{Funcom} selections exhibit similar results. As the dataset size decreases from the \textit{Full} set of 108.7M tokens through SIDE thresholds down to SIDE$_{0.9}$ with 49.0M, the ROUGE scores exhibit slight improvement up to SIDE$_{0.6}$ but remain relatively stable across the golden sets, while METEOR and BLEU-4 scores follow the trend of the ROUGE scores although in a less noticeable way. For \textit{CoderEval Human}, ROUGE scores exhibit minor improvements, while METEOR and BLEU-4 scores show no significant changes. For the Mastropaolo \etal dataset, the effectiveness remains stable across the three metrics and training sizes, further underscoring that increasing the training dataset size has a small impact on performance for high-quality test sets.

Although we observe slight improvements with larger dataset selections (specifically at SIDE$_{0.6}$ and SIDE$_{0.7}$), the gains are minimal, resulting in a generally flat trend on the Pareto front. Finally, as shown in \figref{fig:pareto}, increasing the training dataset size does not lead to substantial improvements in downstream performance.
Based on our results, we can state that increasing dataset size provides negligible improvements, with SIDE$_{0.9}$ achieving an optimal balance between performance and resource efficiency.

\begin{tcolorbox}[
	colframe=black!100!white, colback=black!10!white,
	coltitle=white, colbacktitle=black!,
	title=Answer to RQ$_{1}$,
	fonttitle=\bfseries,
	rounded corners,
	boxrule=0.5mm,
	width=\linewidth,
	breakable
	]
	
	Fine-tuning neural code summarization models with coherent code-comment instances selected through SIDE leads to performances comparable to those obtained when fine-tuning the model on the complete training sets, with a reduction of up to 50\% of the training instances.
\end{tcolorbox}

\begin{table*}[t]
	\centering
	\caption{Performance metrics on Top-1 predictions for \side{0.9} and \textit{Random}.}
	\label{tab:performance_metrics_codereval_comp}
	\resizebox{0.8\linewidth}{!}{%
		\begin{tabular}{c|l|rrr|rrr|rrr}
			\toprule
			\rowcolor{black}
			&  & \multicolumn{3}{c}{\textcolor{white}{\textbf{CoderEval-Original \cite{yu2024codereval}}}} & \multicolumn{3}{c}{\textcolor{white}{\textbf{CoderEval-Human \cite{yu2024codereval}}}} & \multicolumn{3}{c}{\textcolor{white}{\textbf{Mastropaolo \etal \cite{mastropaolo2023robustness}}}} \\
			\rowcolor{gray!20}
			\textbf{Dataset} & \textbf{Selection} & \textbf{BLEU-4} & \textbf{METEOR} & \textbf{ROUGE} & \textbf{BLEU-4} & \textbf{METEOR} & \textbf{ROUGE} & \textbf{BLEU-4} & \textbf{METEOR} & \textbf{ROUGE} \\
			\midrule
			\multirow{2}{*}{\textit{TL-CodeSum \cite{hu2018summarizing}}}
			& \side{0.9} & 11.60 & 17.07 & 33.67 & 6.56 & 14.19 & 30.53 & 5.62 & 12.75 & 27.26   \\
% 			\cmidrule(r){2-11}
			& \emph{Random} & 10.77 & 16.36 & 32.67 & 6.12 & 13.44 & 28.78 & 5.85 & 12.89 & 27.13   \\
			\midrule
			\multirow{2}{*}{\textit{Funcom \cite{leclair2019neural}}}
			& \side{0.9} & 13.65 & 18.08 & 37.12 & 6.78 & 13.61 & 30.07 & 6.81 & 12.95 & 28.07 \\
			& \emph{Random} & 13.67 & 17.95 & 36.09 & 6.88 & 13.95 & 29.90 & 5.84 & 12.32 & 27.12 \\
			\bottomrule
	\end{tabular}}
\end{table*}

\begin{table}[h!]
	\centering
	\caption{Results Comparison for codesum and funcom}
	\label{tab:tests_comp}
	\resizebox{\linewidth}{!}{%
		\begin{tabular}{l|l|r|r|r|r|r|r}
			\toprule
			\rowcolor{black}
			 & & \multicolumn{2}{c}{\textcolor{white}{\textbf{CoderEval-Original \cite{yu2024codereval}}}} & \multicolumn{2}{c}{\textcolor{white}{\textbf{CoderEval-Human \cite{yu2024codereval}}}} & \multicolumn{2}{c}{\textcolor{white}{\textbf{Mastropaolo \etal  \cite{mastropaolo2023robustness}}}} \\
			\rowcolor{gray!20}
			\textbf{Dataset} & \textbf{Metric} & \textbf{$\boldsymbol{p}$-value} & \textbf{Cliff $\boldsymbol{\delta}$} & \textbf{$\boldsymbol{p}$-value} & \textbf{Cliff $\boldsymbol{\delta}$} & \textbf{$\boldsymbol{p}$-value} & \textbf{Cliff $\boldsymbol{\delta}$} \\
			\midrule
			\multirow{3}{*}{\textit{TL-CodeSum \cite{hu2018summarizing}}} & BLEU-4  & 1.0 &   0.03 & 0.5 &   0.03 &   1.0 &   0.02 \\
																		  & METEOR  & 1.0 &   0.02 & 0.5 &  0.014 &   0.4 & -0.013 \\
																		  & ROUGE-L & 1.0 &   0.03 & 0.5 &  0.052 &   1.0 &   0.01 \\
			\midrule
			\multirow{3}{*}{\textit{Funcom \cite{leclair2019neural}}}     & BLEU-4  & 1.0 & -0.004 & 1.0 & -0.021 & 0.017 &  0.042 \\
																	      & METEOR  & 1.0 &   0.01 & 1.0 &  0.014 &  0.11 &  0.025 \\
																	      & ROUGE-L & 1.0 &   0.03 & 1.0 &  0.013 &  0.07 &  0.021 \\
			\bottomrule
		\end{tabular}
	}
\end{table}

\subsection{RQ$_2$. Comparison with the Random Baseline}
\label{rq2}
\tabref{tab:performance_metrics_codereval_comp} reports the performance achieved by the models trained on the \side{0.9} and \textit{Random} selections for both datasets (\ie \textit{TL-CodeSum} and \textit{Funcom}). We report such information as described in \secref{rq1}. Instead, \tabref{tab:tests_comp} reports the results of the Wilcoxon signed-rank tests with the adjusted $p$-values (column \textit{$p$-value}), and the corresponding Cliff's $\delta$ effect sizes (column Cliff's $\delta$) for each golden set, including \textit{CoderEval-Original}, \textit{CoderEval-Human}, and Mastropaolo \etal

The results are quite comparable for the models trained on the \textit{TL-CodeSum} dataset selections. In detail, on \textit{CoderEval-Original}, the model trained on the \side{0.9} selection achieves slightly better results ($\uparrow$ 0.83, $\uparrow$ 0.71, $\uparrow$ 1.0), even though without statistically significant statistical differences ($p$-value = 1.0), and negligible effect sizes for all metrics ($\sim$ 0.00). We obtain consistent results for \textit{CoderEval-Human}. Instead, we observe slight differences in performance for the Mastropaolo \etal golden set. Specifically, the \textit{Random} selection allows to achieve better BLEU-4 ($\uparrow$ 0.23) and METEOR ($\uparrow$ 0.14), but lower ROUGE ($\downarrow$ 0.13). Again, we observe no statistically significant differences and negligible effect sizes.

What was observed above is also generally true for the models trained on the \textit{Funcom}. The model trained on the \side{0.9} selection, achieves better ROUGE ($\uparrow$ 1.03, $\uparrow$ 0.17, $\uparrow$ 0.95) on all the three golden sets. Conversely, for the BLEU and METEOR metrics, the models show alternate better performance, except on the Mastropaolo \etal golden set. \tabref{tab:tests_comp} reports a statistically significant difference between the performance of the two models for the BLEU-4 metric, although with negligible effect sizes.

\begin{tcolorbox}[
	colframe=black!100!white, colback=black!10!white,
	coltitle=white, colbacktitle=black!,
	title=Answer to RQ$_{2}$,
	fonttitle=\bfseries,
	rounded corners,
	boxrule=0.5mm,
	width=\linewidth,
	breakable
	]
	Filtering training instances based on code-comment coherence provides models with comparable effectiveness to those trained on randomly selected instances.
\end{tcolorbox}

\section{Discussions}
\label{sec:implication}
Our findings challenge the conjecture that code-comment coherence, as measured by SIDE \cite{mastropaolo2024evaluating}, is a critical quality attribute for filtering instances of code summarization datasets. By selecting $\langle code, summary \rangle$ pairs with high-coherence for training allow to achieve the same results that would be achieved by randomly selecting such a number of instances. At the same time, we observed that reducing the datasets size up to 50\% of the training instances does not significantly affect the effectiveness of the models, even when the instances are randomly selected. These results have several implications.

First, code-comment consistency might not be a problem in state-of-the-art datasets in the first place, as also suggested in the results of RQ$_0$. Also, the DL models we adopted (and, probably, bigger models as well) are not affected by inconsistent code-comment pairs, even when these inconsistencies are present in the training set.
Despite the theoretical benefits of filtering by SIDE \cite{mastropaolo2024evaluating}, that is the state-of-the-art metric for measuring code-comment alignment, our results indicate its limitations in improving the \textit{overall} quality of the training sets for code summarization task.
Nevertheless, other quality aspects of code and comments that have not been explored yet (such as readability) may be important for smartly selecting the training instances.
Future work should explore such quality aspects further.

Our results clearly indicate that state-of-the-art datasets contain instances that do not contribute to improving the models' effectiveness. This finding is related to a general phenomenon observed in Machine Learning and Deep Learning. Models reach convergence when they are trained for a certain amount of time (epochs). Additional training provides smaller improvements and increases the risk of overfitting. We show that the same is true for data. In terms of effectiveness, model convergence is achieved with fewer training instances than previously assumed. Limiting the number of epochs may make it possible to reach model convergence with a subset of training data, maintaining model effectiveness, reducing resource demands and minimizing the risk of overfitting.
Future work could explore different criteria for data selection that identify the most informative subsets for training.
Conversely, this insight suggests that currently available datasets suffer from poor diversity (thus causing the previously discussed phenomenon).
This latter insight constitutes a clear warning for researchers interested in building code summarization datasets, which should include instances that add relevant information instead of adding more data, which might turn out to be useless.

Finally, it is worth pointing out that another benefit of the reduction we performed is the environmental impact. Reducing the number of training instances implies a reduced training time, which, in turn, lowers the resources necessary to perform training and, thus, energy consumption and CO$_2$ emissions.
We performed a rough estimation of the training time across different selections of \textit{TL-CodeSum} and \textit{Funcom} datasets and estimated a proxy of the CO$_2$ emissions for each model training phase by relying on the ML CO$_2$ impact calculator\footnote{\url{https://mlco2.github.io/impact/\#compute}} \cite{lacoste2019quantifying}. Such a calculator considers factors such as the total training time, the infrastructure used, the carbon efficiency, and the amount of carbon offset purchased. The estimation of CO$_{2}$ emissions needed to train the model with the \textit{Full} selection of \textit{Funcom} ($\sim$ 200 hours) is equal to 26.05 Kg, while with the optimized training set, \ie $SIDE_{0.9}$ ($\sim$ 90 hours), the estimation is 11.69 Kg of $CO_2$ (-55\% emissions).
While we recognize that this method provides an estimation rather than a precise measurement, it offers a glimpse into the environmental impact of applying data reduction.

\section{Threats to Validity}
\label{sec:threats}
% !TEX root = main.tex

This section describes the threats that could affect the findings of our study.

\textbf{Construct validity} threats concern the relationship between theory and observation. A first threat is related to how we assess the quality of the generated summaries. We leverage metrics widely used in literature and, specifically, BLEU-4, METEOR, and ROUGE. We are aware that such metrics may not fully reflect the developers' perception of a summary quality.

Another threat could be related to the choice of SIDE as a driver for selecting a training set based on summary coherence. As shown in previous work \cite{mastropaolo2024evaluating}, this metric correlates with human-based summary evaluation better than other state-of-the-art metrics.

\textbf{Internal validity} threats concern factors internal to our study that could affect our findings. One factor is related to the hyperparameter calibration of the performed training. As explained in \secref{sec:study}, our choices are based on those of previous studies \cite{mastropaolo2023towards,ciniselli2024generalizability}.

Another factor is the choice of the SIDE cut thresholds. We have chosen five thresholds varying from 0.5 to 0.9 and reported the findings for such levels of dataset filtering. While we did not explore the full range of SIDE values, we have evaluated, with a discretization of 0.1, the whole range from 0.5 above. We conjecture that lower values would not show results much different from the full datasets because the 0.5 threshold indicates a very small reduction (7\%-9\% on the two datasets).

Finally, we are aware that a more accurate computation of training times would require multiple runs. However, this was unfeasible given the number of configurations to evaluate and the training time required for each of them. 

\textbf{Conclusion validity} threats concern the relationship between the experimentation and outcome. The study is mostly observational. Therefore, we report and discuss results through descriptive statistics. However, wherever appropriate---and in particular for RQ$_1$ and RQ$_2$, we complement them with suitable statistical procedures (Wilcoxon rank-sum test and Cliff's $d$ effect size). Also, since multiple tests have been performed, we adjust the $p$-values using Holm's correction procedure \cite{holm1979simple}.

\textbf{External validity} threats concern the generalizability of our findings. We have considered two datasets for the training. These datasets have been specifically designed for code summarization and cleaned up using the CAT approach by Shi \etal \cite{shi2022we}. 
As for the test set, we have considered two datasets that (i) do not overlap with the training set and (ii) have been evaluated by humans. 
Although these datasets are particularly suitable for the study reported in this paper, we cannot exclude that applying the selection strategy to other datasets, particularly those related to different programming languages or application domains, might lead to different results.

\section{Related Work}
\label{sec:related}
% !TEX root = main.tex
In this section, we examine key literature concerning the impact of data quality on neural models for automating software engineering practices, with a particular focus on code summarization.
Following this, we provide a section covering the main approaches in code summarization literature.

\subsection{The Importance of Data Quality for DL-Based Methods in Code Summarization}
\label{sec:dataset-cleaning}

Data quality is essential for the success of DL methods designed to automate software engineering tasks. Models trained on noisy data frequently suffer significant performance degradation when deployed in real-world settings. A recent study by Shi \etal \cite{shi2022evaluation} examines various factors that could impact the evaluation of code summarizers, including metrics, code preprocessing operations, and datasets. Their empirical study showed that these elements significantly influence the evaluation of code summarization models. Specifically, Shi \etal observed that certain code preprocessing operations (\eg converting all tokens to lowercase) could substantially affect performance. This can either enhance the model's effectiveness or lead to a decline in performance, showcasing the fundamental role of preprocessing strategies to either boost or impair model outcomes. 

LeClair and McMillan \cite{leclair2019recommendations} also explored the effects of different preprocessing decisions on datasets used for code summarization. Specifically, they analyzed how splitting training and test datasets by function versus by project influences model performance.

Gros \etal \cite{gros2020code} investigate the relationship between code comments and code itself, particularly exploring the premise that generating code comments is akin to translating between natural languages. This conceptual alignment has facilitated using models and evaluation metrics from Natural Language Processing (NLP) in tasks like code summarization. Gros \etal assessed how code-comment datasets, 
compare with datasets used for natural language translation. Specifically, they contrasted each code-comment dataset against WMT19~\cite{barrault2019findings}, a renowned corpus utilized for training natural language translators. The outcomes of their research indicated that code comments are more repetitive than English sentences found in natural language, a characteristic that influences performance evaluation metrics.

Sun \etal \cite{sun2022importance} illustrate the issue of poor natural language data quality within the context of code search—specifically, retrieving the ``closest'' code matching a natural language query in a codebase. By implementing a semantic query cleaning module for code search datasets, Sun \etal showed that their filtering framework not only enhances the accuracy of models trained on the filtered dataset but also conserves computing resources. Along these lines, Li \etal~\cite{li2023commit} investigated the quality of commit messages and found that (i) the quality of commit messages plays a fundamental role when it comes to software defects, and (ii) the overall quality of these messages declines over time, despite the developers' belief who think their commit messages are improving.

Xu \etal \cite{xu2023data} explored the impact of data quality on just-in-time obsolete comment detection.
They empirically showed that applying a set of manually derived rules could enhance accuracy by up to 10.7\%.

Shi \etal \cite{shi2022we} implemented a similar approach to address data quality issues in code summarization benchmarks by introducing CAT (Code-comment Cleaning Tool). This tool can identify noisy data in various programming languages, including Java and Python. Specifically, the rule-based method (\ie CAT) developed by Shi \etal is designed to detect specific patterns of noisy data at both the comment and code levels, based on a taxonomy of data preprocessing noises identified across four popular datasets. An initial evaluation with CAT showed that commonly used code summarization benchmark datasets for Python and Java include noisy data pairs $\langle code, comment \rangle$, ranging from 31\% to 66\%. These noisy elements were either ``fixed'' to decrease their noise level or completely removed. Following this cleanup, state-of-the-art neural code summarization techniques were retrained from the ground on these refined benchmarks. The performances of the different models were then compared against the cleaned test datasets processed using CAT. This comparison indicated that the optimized training dataset significantly improved summarization accuracy and overall performance.

\subsection{Automated Code Summarization}
Different studies have explored the automation of code summarization, with three primary approaches emerging in the literature: Information Retrieval (IR), Deep Learning (DL), and hybrid methods combining both techniques.

One of the first approaches to DL-based code summarization is the work by Iyer \etal \cite{iyer2016summarizing}, who introduced an RNN-based model with an attention mechanism for generating code summaries. They used an encoder-decoder architecture, proposing CODE-NN to generate summaries. It was trained on code-description pairs from StackOverflow and it demonstrated substantial improvements over traditional approaches in generating summaries, highlighting one of the first evidence of DL effectiveness for code summarization tasks.

Zhang \etal \cite{zhang2020retrieval} proposed a retrieval-based neural model, Rencos, which combines information retrieval (IR) techniques with neural machine translation (NMT) specifically for code summarization. They first train an encoder-decoder model. Then, during the testing phase, they retrieve the most similar snippets from the training set in terms of syntax and semantics and encode them with the input. Finally, after fusing them it predicts the summary.

One of the first works exploring Transformer-based approaches for code summarization is the one by Ahmad \etal \cite{ahmad2020transformer}. In their work, they leverage the self-attention mechanism within Transformers to model the complex, long-distance dependencies in source code, aiming to generate appropriate summaries. Their results demonstrated that Transformers are effective for code summarization tasks.

With the advent of Large Language Models (LLMs), substantial progress has been made in automating code summarization. LLMs are capable of few-shot learning \ie providing task-specific examples in the prompt, allowing the model to perform the requested task.
As first steps for LLM-based code summarization, Ahmed \etal \cite{ahmed2024automatic} introduced the Automatic Semantic Augmentation of Prompts (ASAP) technique. This latter enhances LLMs performance by adding structured semantic information in the prompts. ASAP inserts (i) repository context, (ii) tagged identifiers, and (iii) data flow graphs directly into the prompt, guiding the model toward a wider knowledge of the code structure and functionality. Such technique allowed models like Code-Davinci and GPT-3.5 to achieve state-of-the-art performance on multiple programming languages for the code summarization task.

As part of advancing LLM-based code summarization, Sun \etal \cite{sun2024source} conducted a study examining various prompting strategies and model configurations for improving performance in code summarization tasks. They evaluated prompting techniques such as zero-shot, few-shot, and chain-of-thought, finding that simpler approaches often performed as well as more complex methods.

\subsection{Code-comment Coherence}
Detecting code comments that result to be inconsistent with respect to the given code has been the subject of several works.
Tan \etal \cite{tan2007icomment} proposed \textit{iComment}, a tool that leverages NLP properties, machine learning models, and program analysis techniques to extract rules for a decision tree classifier able to detect potential inconsistencies between comments and code. They evaluated their tool for large-scale projects such as Linux, Mozilla, Apache, and Wine, finding several inconsistencies, including bugs and bad comments.

Liu \etal \cite{liu2018automatic}, instead, focused on detecting outdated comments during code changes.
The authors proposed a machine learning-based approach that leverages 64 features related to code, comments, and their relationship before and after code changes to identify outdated comments. Using a random forest classifier, they evaluated their approach on several open-source projects and demonstrated its effectiveness in detecting outdated comments, achieving high precision and recall rates.

Wang \etal \cite{wang2019deep} proposed DComment, a framework to evaluate the quality of comments in source code. By analyzing both the code and its associated comments, DComment identifies meaningful patterns and relationships to determine whether a comment is coherent and relevant. DComment showed a strong ability to generalize across different projects on multiple datasets.

Xu \etal \cite{xu2024code} proposed MCCL, which combines method comment detection and confidence learning to identify and mitigate inconsistencies. The proposed approach uses advanced encoding techniques, including multi-head attention and graph neural networks, to analyze the relationships between code and comments. Additionally, a denoising component addresses noisy data and labeling errors, improving the detection process. MCCL was evaluated on over 1,500 open-source projects, demonstrating superior performance compared to existing methods, significantly improving precision, recall, and F1-score.

In this work, we adopt SIDE \cite{mastropaolo2024evaluating} as an approach for measuring code-comment coherence instead of the previously-mentioned approaches because, differently from them, (i) it has been shown to highly correlate with human judgment, and (ii) it allows us to filter instances based on a threshold (multiple selectivity levels).

\section{\rev{Conclusion}}
\label{sec:conclusion}
% !TEX root = main.tex

We presented an empirical investigation in which we studied how filtering out incoherent code-comment pairs in code summarization datasets affects the DL models trained to tackle this task in terms of effectiveness and removed training instances. 

On the one hand, our results show that reducing the number of instances in the training set using a coherence-based approach driven by the SIDE metric \cite{mastropaolo2024evaluating} does not significantly impact the models' effectiveness, even when applying the strictest filter that reduces the training set of over 50\%. On the other hand, we found that randomly selecting the same number of instances instead of selecting them based on their code-comment coherence allows us to achieve comparable results. 

Our future research agenda includes investigating other quality aspects for selecting training data, such as instance diversity and readability. Furthermore, we plan to investigate how far we can go in randomly removing instances without affecting the quality of generated code summaries.

\section{Data Availability}
\label{sec:data}
The study dataset and scripts used for the analysis are available and documented in our online replication package~\cite{replicationpackage}.

\section{Acknowledgments}
\label{sec:ack}
This publication is part of the project PNRR-NGEU which has received funding from the MUR – DM 118/2023.
This work has been partially supported by the European Union - NextGenerationEU through the Italian
Ministry of University and Research, Projects PRIN 2022 ``QualAI: Continuous Quality Improvement of AI-based Systems'', grant n. 2022B3BP5S, CUP: H53D23003510006.

\balance
\bibliographystyle{IEEEtran}
\bibliography{main}

\end{document}